\documentclass[11pt]{amsart}
\usepackage{amssymb}
\usepackage{latexsym}
\usepackage{graphicx}
\usepackage{tikz}
\usepackage[linktocpage=true]{hyperref}
\usepackage{ulem}
\usepackage[left=3.8cm, right=3.8cm, top=3cm, bottom=3cm]{geometry}
\usepackage{color}

\hypersetup{ colorlinks   = true, 
urlcolor  = blue, 
linkcolor    = blue, 
citecolor   = blue 
}

\def\beq{\begin{equation}}
\def\eeq{\end{equation}}

\newcommand{\Z}{{\mathbb Z}}
\newcommand{\R}{{\mathbb R}}

\newcommand{\C}{{\mathbb C}}

\newcommand{\T}{{\mathbb T}}

\newcommand{\N}{{\mathbb N}}

\newtheorem{Theorem}{Theorem}[section]
\newtheorem{Corollary}[Theorem]{Corollary}
\newtheorem{Proposition}[Theorem]{Proposition}
\newtheorem{Lemma}[Theorem]{Lemma}
\newtheorem{lemma}[Theorem]{Lemma}

\newtheorem{definition}[Theorem]{Definition}

\newtheorem{corollary}{Corollary}

\newcommand{\la}{\langle}
\newcommand{\ra}{\rangle}

\def\lkr#1{\langle #1\rangle_{\omega}}

\def\im{{\rm i}}

\def\cI{{\mathcal{I}}}
\def\cK{{\mathcal{K}}}
\def\cS{{\mathcal{S}}}
\def\cC{{\mathcal{C}}}
\def\cJ{{\mathcal{J}}}
\def\cL{{\mathcal{L}}}
\def\arcos{{\rm arcos}}

\begin{document}

\title[]{Dispersive estimate for quasi-periodic Schr\"odinger operators on 1-$d$ lattices}

\author{Dario Bambusi}
\address{Dipartimento di Matematica, Universit\`a degli Studi di Milano, Milano, Italy}
\email{dario.bambusi@unimi.it}

\author{Zhiyan Zhao}
\address{Laboratoire J.A. Dieudonn\'{e}, Universit\'e C\^ote d'Azur, 06108 Cedex 02 Nice, France}
\email{zhiyan.zhao@univ-cotedazur.fr}
\thanks{Research of Z. Zhao is partially supported by ANR grant ``ANR-15-CE40-0001-03" BeKAM.
Z. Zhao is also partially supported by the French government, through the UCA JEDI Investments in the Future project managed by the National Research Agency (ANR) with the reference
number ANR-15-IDEX-01.}

\begin{abstract}
Consider the one-dimensional discrete Schr\"odinger operator $H_{\theta}$:
$$(H_{\theta} q)_n=-(q_{n+1}+q_{n-1})+ V(\theta+n\omega) q_n \ , \quad n\in\Z \ ,$$
with $\omega\in\R^d$ Diophantine, and $V$ a real-analytic function on $\T^d=(\R/2\pi\Z)^d$.
For $V$ sufficiently small, we prove the dispersive estimate: {for every $\phi\in\ell^1(\Z)$,}
\begin{equation}
  \label{dispersive}
  \|e^{-{\rm i}tH_{\theta}}\phi\|_{\ell^\infty} \leq K_0  \frac{
    |\ln\varepsilon_0|^{a(\ln\ln(2+\langle t\rangle))^2 d}} {\langle
    t\rangle^{\frac13}} \|\phi\|_{\ell^1} \ , \quad \la t \ra:=\sqrt{1+t^2} \ ,
\end{equation}
{with $a$ and $K_0$ two absolute constants} and
$\varepsilon_0$ an analytic norm of $V$. The estimate holds for every
$\theta\in\T^d$.
\end{abstract}

\maketitle


\section{Introduction and main results}

\noindent

 Consider the quasi-periodic Schr\"odinger operator $H_{\theta}:\ell^2(\Z)\to\ell^2(\Z)$, defined as
\begin{equation}\label{qpoperator}
(H_{\theta}q)_n=-(q_{n+1}+q_{n-1})+V(\theta+n\omega)q_n \ , \quad n\in\Z \ ,
\end{equation}
with $V:\T^d\to\R$ an analytic potential, $d\geq 1$, and $\omega\in \R^d$ a
Diophantine frequency vector,  it is well
known that its spectrum, that we shall denote by
$\Sigma$, is independent of $\theta$.  It is also well known that when
the potential function $V$ is sufficiently small, the operator
$H_{\theta}$ has purely absolutely continuous spectrum (see e.g. \cite{A,
  E92}, see also \cite{AD}) and that for generic potential it is a Cantor set. Furthermore, the
time evolution $e^{-{\rm i}tH_\theta}$ presents ballistic transport
(see \cite{Zhao}).

In the present paper we prove that $e^{-{\rm i}tH_\theta}$ also
fulfills the $\ell^1$-$\ell^\infty$ dispersive estimate
\eqref{dispersive}. As usual, from this estimate one can deduce
Strichartz estimates \cite{KT} as well as decay and scattering for the small
amplitude solutions of the nonlinear Schr\"odinger equation
\begin{equation}
  \label{nls}
\im \dot q_n=(H_\theta q)_n\pm\left|q_n\right|^{p-1} q_n  \ , \quad n\in\Z  \ ,
\end{equation}
provided $p$ is large enough {(see e.g. \cite{SK, KPS})}. Here we
concentrate just on initial data in $\ell^1$ and dispersive decay in
$\ell^\infty$ and give the result for $p>5$.

We recall that for the free Schr\"odinger operator,
\begin{equation}
  \label{delta}
(-\Delta q)_n:=-(q_{n+1}+q_{n-1}) \ , \quad n\in\Z \ ,
\end{equation}
the $\ell^1$-$\ell^{\infty}$ estimate
\begin{equation}\label{disp_free}
\|e^{{\rm i}t\Delta}\phi\|_{\ell^\infty} \leq \frac{C}{\langle
t\rangle^{\frac13}}  \| \phi\|_{\ell^1} \ , \quad \forall
\ \phi\in\ell^1(\Z) \ , \quad \langle t\rangle:=\sqrt{1+t^2} \ ,
\end{equation}
is well known (see \cite{SK}, see also \cite{MP10}).
For the operator
$H:\ell^2(\Z)\to\ell^2(\Z)$,
$$(Hq)_n=-(q_{n+1}+q_{n-1})+V_n q_n \ , \quad n\in\Z \ ,$$
Pelinovsky-Stefanov \cite{PS} have shown that
\begin{equation}\label{disp_pointdecay}
\|e^{-{\rm i}tH}P_{ac}\phi\|_{\ell^\infty} \leq \frac{C}{\langle t\rangle^{\frac13}}\|\phi\|_{\ell^1} \ , \quad \forall \ \phi\in\ell^1(\Z) \ ,
\end{equation}
for ``generic"\footnote{See Definition 1 of \cite{PS}.} potentials
$V_n$ decaying sufficiently fast at infinity. Here $P_{ac}$ denotes the
projection on the absolutely continuous part of the spectrum.
{For other related works, one can refer to \cite{KKK, KPS, CucTar09, bam13a, EKT}.}

In all these examples the continuous spectrum is the union of disjoint
intervals. We emphasize that our result is the first one in which the
continuous spectrum is a Cantor set.

In order to state precisely our main theorem we need a few
preliminaries.

\begin{definition}
  \label{dio.1}
A vector $\omega\in\R^d$ will be said to be {\it Diophantine} if
$\exists \ \gamma>0$ and $\tau> d-1$, s.t,
\begin{equation}\label{dio}
\inf_{j\in\Z}\left|\la k, \omega\ra-j\pi\right| > \frac{\gamma}{|k|^{\tau}}\ , \quad \forall  \  k\in\Z^d\setminus\{0\} \ ,
\end{equation}
where $\la \cdot,\cdot \ra$ is the scalar product on $\R^d$.
\end{definition}
We will assume that there exists a positive $r$ s.t. the potential
extends to a bounded complex analytic function on $|\Im\theta|<r$. We
will denote
\begin{equation}
  \label{vr}
\varepsilon_0:=\left|V\right|_r:=\sup_{|\Im\theta|<r}\left|V(\theta)\right| \ .
\end{equation}
We will also denote this class of functions by $\cC^\omega_r(\T^d)$.

\smallskip

Our main result is the following theorem.

\begin{Theorem}\label{dispersion_log}
There exists $\varepsilon_{*}=\varepsilon_{*}(r, \gamma,\tau,d)>0$ and
two absolute constants $a$, $K_0>0$ such that if $\varepsilon_0<
\varepsilon_{*}$, then for any $\theta\in\T^d$, any $t\in\R$, the following estimate holds: \begin{equation}
  \label{main}
\|e^{-{\rm i}tH_{\theta}}\phi\|_{\ell^\infty} \leq  K_0 \frac{
    |\ln\varepsilon_0|^{a(\ln\ln(2+\langle t\rangle))^2 d}} {\langle
    t\rangle^{\frac13}}
\|\phi\|_{\ell^1} \ , \quad \forall \ \phi\in\ell^1(\Z) \ .
\end{equation}
\end{Theorem}
It is immediate to get the following

\begin{corollary}\label{cor.pot}
Assume $\varepsilon_0<\varepsilon_*$, with $\varepsilon_*$ as
    in Theorem \ref{dispersion_log}, then given any $0<\zeta<\frac13$,
    there exists $K_1=K_1(\varepsilon_0,\zeta)$ s.t. for any
    $\theta\in\T^d$, any $t\in\R$,
\begin{equation}
  \label{linest}
\|e^{-{\rm i}tH_{\theta}}\phi\|_{\ell^\infty} \leq  \frac{ K_1
} {\langle
    t\rangle^{\zeta}}
\|\phi\|_{\ell^1} \ , \quad \forall \ \phi\in\ell^1(\Z) \ .
\end{equation}
\end{corollary}

One also has the following standard corollary on the $\ell^\infty$
decay of the solution of \eqref{nls} with $p>5$ and small $\ell^1$
initial datum.

\begin{corollary}
  \label{NLS}
Consider Eq. \eqref{nls} with $p>5$, assume
    $\varepsilon_0<\varepsilon_{*}$ and fix $\zeta$ fulfilling
$$\frac{1}{p-2}<\zeta<\frac{1}{3} \ .$$ Then there exists
    $\delta_*>0$, with $\delta_*=\delta_*(r,
    \gamma,\tau,d,\epsilon_0,\zeta)$ such that if the initial datum
    $\phi=q(0)$ fulfills
$$
\delta_0:=\left\|\phi\right\|_{\ell^1(\Z)}< \delta_* \ ,
$$
then the solution $q(t)$ of \eqref{nls} fulfills
\begin{equation}
  \label{nls.esti}
\left\|q(t)\right\|_{\ell^{\infty}}\leq \frac{4K_1}{\langle
  t\rangle^{\zeta}}\left\|\phi\right\|_{\ell^1(\Z)}\ ,
  \end{equation}
  where $K_1$ is the constant in Corollary \ref{cor.pot}.
\end{corollary}
For the sake of completeness, we will give the proof of Corollary \ref{NLS} in Section \ref{nonlinear}.

\smallskip

From \eqref{main} one can also deduce, as in \cite{SK, KPS},
  Strichartz estimates as well as decay and scattering for all the
  solutions of the linear Schr\"odinger equation. From this one can
  also deduce scattering for all solutions of \eqref{nls} with small initial
  data in the energy space $\ell^2$, {\it provided $p>7$}.

\

\noindent {\it Scheme of the proof of Theorem \ref{dispersion_log}.}
For the free Schr\"odinger operator \eqref{delta}, the dispersive estimate
is proved by using the Fourier transform which allows to write
$e^{{\rm i}t\Delta}\phi$ as an oscillatory
integral, which is estimated through the Van der Corput lemma which
gives the $t^{-\frac13}$ decay. The variable of integration in the
integral to be estimated is the wave number.

In the presence of a quasi-periodic potential, generically, the spectrum
is a Cantor set and the object generalizing the wave number is the
fibered rotation number of the corresponding
Schr\"odinger cocycle (see Appendix \ref{rotation} for a precise
definition).

Now, in the quasi-periodic case the fibered rotation number (rotation
number for short) can be approximated through a perturbative
construction. After $J$ steps of such a construction, the
approximate rotation number $\rho_J$ is a monotonic function of class
$\cC^k$ (in our case $k=3$ is enough), which is defined on the union
of a very large number of intervals, precisely there is a total number of
intervals proportional to $\left|\ln\varepsilon_0\right|^{2J^2d}$,
but the approximate rotation number behaves as $E^{\frac12}$ at
the boundaries of each interval ($E$ being the spectral parameter), so
that its derivatives diverges at such points, which in the limit are
dense in the spectrum.

Following \cite{Zhao}, the idea of the proof is to stop the
construction at some step, say the $J$-th one, and to apply Van
Der Corput lemma on each one of the small intervals.
Still one has to make a regularization at the boundaries of the intervals, and this will be explained in a while.
First, one has to know the
improper eigenfunctions of $H_{\theta}$. Now, it is known how to
construct such improper eigenfunctions perturbatively: they are the
quasi-periodic solutions of the quasi-periodic cocycle associated to
$H_\theta$. However, it is not known how to construct the spectral
measure and how to normalize the improper eigenfunctions. The idea is
to choose an approximate normalization, which in some sense is the
most natural one, and to modify it slightly in order to regularize
the integrals to be estimated. It turns out that this is possible, and
that, if one uses such ``normalized'' eigenfunctions to define a
spectral transform, then such a transformation is not unitary, but
it is bounded with a bounded inverse and thus suffices to get the result.
This was done and proved in \cite{Zhao}. Here we just recall the needed results.

In the present paper we use such a spectral transform in order to
write down an approximate representation formula for the solution of
the Schr\"odinger equation in terms of oscillatory integrals that we
estimate by approximating them through integrals over intervals which
in turn are estimated through the Van Der Corput Lemma.  In order to
get the result, the last difficulty is to estimate the errors related
to the use of approximations. This is purely technical and consists in
writing down all the estimates taking into account the dependence on
the approximation step and on the other parameters and then to choose
all the free parameters in a suitable way. The main technical lemma of
the paper gives this estimate and is Lemma \ref{h_integral_t}.

\smallskip

The rest of paper is organized as follows. In
  Sect. \ref{sec_pre_notation} we recall some known facts on the structure
  of the spectrum of the Schr\"odinger operator and on the
  construction of the spectral transform. In Sect. \ref{sec_int} we
  prove the main technical lemma of the paper, namely Lemma
  \ref{h_integral_t}. In Sect. \ref{sec_proof} we conclude the proof
  of the main theorem. In Sect. \ref{nonlinear} we prove
    Corollary \ref{NLS}. We also add two Appendixes. In Appendix
  \ref{rotation} we recall a few facts on the rotation number, while
  in Appendix \ref{corput} we recall the version of the Van Der
  Corput Lemma that we use in the paper.

\section{Preliminaries on Schr\"{o}dinger operator and Schr\"{o}dinger
  cocycle}
\label{sec_pre_notation}

 In this  section, we recall some basic notions and some important
 results for the spectrum of the quasi-periodic Schr\"{o}dinger operator
 $H_\theta:\ell^2(\Z)\rightarrow \ell^2(\Z)$,
$$
(H_\theta q)_n=-(q_{n+1}+q_{n-1})+V(\theta+n\omega) q_n \ , \quad n\in\Z \ ,
$$
with $V$ and $\omega$ given as in the statement of Theorem \ref{dispersion_log}.
We will also consider the Schr\"{o}dinger cocycle $(\omega, A_0+F_0)$:
\begin{equation}\label{qpcocycle}
\left(\begin{array}{c}
q_{n+1} \\
q_{n}
\end{array}
\right)=(A_0(E)+F_0(\theta+n\omega))\left(
\begin{matrix}
q_{n} \\
q_{n-1}
\end{matrix}
\right) \ ,
\end{equation}
with $A_0(E):=\left(\begin{array}{cc}
            -E & -1 \\
            1 & 0
          \end{array}
\right)$ and $F_0(\cdot):=\left(\begin{array}{cc}
            V(\cdot) & 0 \\
            0 & 0
          \end{array}
\right)$. Note that $(\omega, A_0+F_0)$ is equivalent to the eigenvalue problem $H_\theta q=Eq$.

\subsection{Structure of the spectrum}\label{spectrum}
\noindent

We review here the KAM theory of Eliasson \cite{E92} and Hadj Amor
\cite{HA} for the reducibility of the Schr\"odinger cocycle $(\omega,
A_0+F_0(\cdot))$. These works relate the reducibility and the fibered
rotation number (for the definition see Appendix \ref{rotation})
globally, and improve the previous works by Dinaburg-Sinai
\cite{DiSi} and Moser-P\"oschel \cite{MP}. Here we will not prove the
corresponding results (Theorem \ref{propsana} and \ref{propsana1})
referring to the work \cite{Zhao} where a detailed proof was
given. However, we will explain the strategy of proof with the aim of
making the paper as self contained as possible, without adding too
many details on known facts.

With $\varepsilon_0=|V|_r$, $\sigma=\frac{1}{200}$, define, as in
\cite{HA}, the sequences:
$$\varepsilon_{j+1}=\varepsilon_{j}^{1+\sigma} \ , \;\
N_j=4^{j+1}\sigma|\ln\varepsilon_j| \ , \quad j\geq 0 \ .$$
All along the paper we will denote
\begin{equation}
  \label{stor}
\lkr k:=\frac{\langle k,\omega\rangle}{2} \ , \quad  k\in\Z^d \ ,
\end{equation}
and by $\left|\cdot \right|_{\cC_W^{k}(\cS)}$ the $\cC^k$
norm of a function which is Whitney smooth on a set $\cS\subset\R$, and for a function which is analytic on $\T^d$ (or $2\T^d$) and Whitney smooth on $\cS$, we will denote by $\left|\cdot \right|_{\cC_W^{k}(\cS), \T^d}$ or $\left|\cdot \right|_{\cC_W^{k}(\cS), 2\T^d}$ the supremum norm on $\T^d$ (or $2\T^d$) and $\cC_W^k$ norm on $\cS$. In particular, if $\cS$ is a union of finitely many intervals, we will omit the subscript $W$ in the above norms.

Furthermore, we denote the fibered rotation number of the Schr\"odinger cocycle
$(\omega,A_0+F_0)$ by $\rho\equiv\rho_{(\omega,A_0+F_0)}$.
It is necessary to mention that $\rho:\R\to [0,\pi]$ is a non-decreasing function with
$$\rho(E)\left\{ \begin{array}{ll}
=0 \ , & E\leq \inf\Sigma  \\
\in (0,\pi) \ ,& E\in (\inf\Sigma, \sup\Sigma) \\
=\pi \ ,& E\geq \sup\Sigma
\end{array}  \right.  \ , $$
By the gap-labeling theorem \cite{JM}, $\rho$ is constant in a gap of $\Sigma$  (i.e., an interval on $\R$ in the resolvent set of $H_\theta$), and each gap is labeled with
$k\in \Z^d$ such that $\rho=\lkr k$ mod $\pi$ in this gap.

\begin{Theorem} \label{propsana} There exists $\varepsilon_*=\varepsilon_*(\gamma,\tau, r,d)>0$ such that if $|V|_r=\varepsilon_0<\varepsilon_*$,
 then, for any $j\in \N$, there exists a Borel set $\Sigma_{j}\subset
 \Sigma$, with $\{\Sigma_{j}\}_j$ mutually disjoint, satisfying
\begin{align*}
|\rho\left(\Sigma_{j+1}\right)|&\leq 3 |\ln\varepsilon_j|^{2 d}
\varepsilon_{j}^{\sigma} \ , \quad j\geq 0 \ ,
\\ \left|\Sigma\setminus\widetilde \Sigma\right|&=0 \ ,\quad \widetilde
\Sigma:=\cup_{j\geq 0}\Sigma_{j}
\end{align*}
such that the following statements hold.
 \begin{itemize}
\item [(1)] The Schr\"odinger cocycle $(\omega, A_0+F_0)$ is {\it
  reducible} on $\widetilde
\Sigma$. More precisely, there
  exist $Z$ and $B$, with $Z:\widetilde
\Sigma \times2\T^d\to SL(2,\R)$
  analytic on $2\T^d$ and $B:\widetilde
\Sigma\to SL(2,\R)$
  s.t. $Z$ conjugates $A_0+F_0$ to $B$, namely
$$Z(\cdot+\omega)^{-1} (A_0+F_0(\cdot)) \, Z(\cdot)=B \ .$$
Furthermore
$B$ is ${\cC}^1$ in the sense of Whitney on each ${\Sigma}_j$, and
\begin{equation}\label{limit_state_whitney}
|B-A_0|_{{\cC}^1_W({\Sigma}_0)}\leq \varepsilon_0^{\frac13} \ ; \qquad |B|_{{\cC}^1_W({\Sigma}_{j+1})}\leq N_j^{10\tau} \ ,\quad j\geq 0 \ .
\end{equation}
\item [(2)] The eigenvalues of $B\big|_{\Sigma_j}$, are of the form
  $e^{\pm{\rm i}\xi}$, with $\xi\in \R$, and, for every $j\geq 0$, there is $k_j:\widetilde
\Sigma\rightarrow\Z^d$, such that
\begin{itemize}
  \item[$\bullet$]$0<|k_j|\leq N_{j}$ on $\Sigma_{j+1}$, and $k_l=0$
    on $\Sigma_j$ for $l\geq j$,
  \item[$\bullet$]  $\xi=\rho-\sum_{l\geq0} \lkr{k_l}$ and $0<|\xi|_{\Sigma_{j+1}}< 2 \varepsilon_{j}^{\sigma}$.
\end{itemize}
 \end{itemize}
\end{Theorem}

Theorem \ref{propsana} describes the result of a KAM procedure. If
one stops the procedure at a finite step one gets a picture that will
be needed for our construction and which is contained in the next
theorem (which of course constitutes the main step for the proof of
Theorem \ref{propsana}).

\begin{Theorem}\label{propsana1}
Let $|V|_r=\varepsilon_0< \varepsilon_*$ be as in Theorem \ref{propsana}.
Given any $J\in\N$, for $0\leq j\leq J$, there exists $\Gamma^{(J)}_j\subset[\inf\Sigma, \sup\Sigma]$, satisfying
\begin{itemize}
\item $\Sigma_{j}\subset \Gamma^{(J)}_{j}$ for $0\leq j\leq J$,
\item $\{\Gamma^{(J)}_j\}_{j=0}^{J}$ are mutually disjoint and $\overline{\bigcup_{j=0}^{J}\Gamma^{(J)}_j}=[\inf\Sigma,\sup\Sigma]$
\item $\bigcup_{j=0}^J \Gamma^{(J)}_j$ consists of at most
$|\ln\varepsilon_0|^{2 J^2 d}$ open intervals,
\item If $J\geq 1$, then
$
\left|\rho\left(\Gamma^{(J)}_{j+1}\right)\right|\leq 3|\ln\varepsilon_j|^{2 d} \varepsilon_{j}^{\sigma}
$ for $0\leq j \leq J-1$.
\end{itemize}
Furthermore, the following statements hold.

\noindent
{\bf (S1)}
There exist
$\left\{ \begin{array}{l}
            A_J:\Gamma_j^{(J)}\rightarrow SL(2,\R)\\[1mm]
            F_J:\Gamma_j^{(J)}\times \T^d\rightarrow gl(2,\R) \ analytic \  on \  \T^d\\[1mm]
            Z_J:\Gamma_j^{(J)}\times2\T^d\rightarrow SL(2,\R) \ analytic \  on \ 2\T^d
          \end{array}
\right.$, $0\leq j \leq J$,

\noindent all of which are smooth on each connected component of $\Gamma^{(J)}_j$,
such that
$$Z_J(\cdot+\omega)^{-1} (A_0+F_0(\cdot)) \, Z_J(\cdot)=A_J+F_J(\cdot) \ ,$$
with {$|F_J|_{{\cC}^3(\Gamma^{(J)}_{j}),\T^d}\leq \varepsilon_J$}, $0\leq j \leq J$, and
\begin{equation}\label{sigma_m_0}
|A_{J}- A_0|_{{\cC}^3(\Gamma^{(J)}_{0})}\leq \varepsilon_0^{\frac12} \ , \qquad
|Z_{J}-Id.|_{{\cC}^3(\Gamma^{(J)}_{0}),2\T^d}\leq \varepsilon_0^{\frac13} \ .
\end{equation}
If $J\geq 1$, then for $0\leq j\leq J-1$,
\begin{equation}\label{sigma_m_J}
|A_{J}|_{{\cC}^3(\Gamma^{(J)}_{j+1})}\leq \varepsilon_j^{-\frac{\sigma}6} \ , \qquad
|Z_{J}|_{{\cC}^3(\Gamma^{(J)}_{j+1}),2\T^d}\leq \varepsilon_j^{-\frac{\sigma}3} \ ,
\end{equation}
and, on $\Gamma^{(J)}_{j+1}$,
\begin{equation}\label{trace_A_J}
\varepsilon_{j}^{\frac{\sigma}{4}}\leq |({\rm tr}A_J)'| \leq N_{j}^{10\tau} \ .
\end{equation}
Moreover, for $0\leq j\leq J$,
\begin{equation}\label{error_whitney}
|A_{J}- B|_{{\cC}^1_W(\Sigma_j)} \leq \varepsilon_{J}^{\frac14} \ ,\quad |Z_J-Z|_{{\cC}^1_W(\Sigma_j), 2\T^d}\leq \varepsilon_{J}^{\frac14} \ .
\end{equation}

 \smallskip

\noindent
{\bf (S2)} $A_{J}$ has two eigenvalues $e^{\pm{\rm i}\alpha_{J}}$ with $\alpha_J\in \R \cup {\rm i}\R$. For $\xi_{J}:=\Re\alpha_{J}$, we have
\begin{itemize}
\item $|\xi_{J}-\xi|_{\Sigma_j}\leq \varepsilon_{J}^{\frac14}$, $0\leq j\leq J$.
\item $|\xi_{J}-\rho|_{\Gamma^{(J)}_{0}}\leq \varepsilon_J^{\frac14}$.
\item If $J\geq 1$, then
\begin{itemize}
  \item $|\xi_{J}|_{\Gamma^{(J)}_{j+1}}\leq \frac32\varepsilon_j^{\sigma}$, $0\leq j\leq J-1$.
  \item There is $k_j: \bigcup_{l=0}^{J}\Gamma^{(J)}_l\rightarrow\Z^d$, $0\leq j\leq J-1$, constant on each connected component of $\bigcup_{l=0}^{J}\Gamma^{(J)}_l$, with $0<|k_j|\leq N_{j}$ on $\Gamma^{(J)}_{j+1}$ and $k_l=0$ on $\Gamma^{(J)}_{j+1}$ for $l\geq j+1$ such that
    $\left|\xi_{J} + \sum_{l=0}^{J-1}\lkr{ k_l}-\rho\right|_{\Gamma^{(J)}_{j+1}}\leq \varepsilon_J^{\frac14}$.
\end{itemize}
\end{itemize}

 \smallskip

\noindent
{\bf (S3)} $\bigcup_{j=0}^J\{\Gamma^{(J)}_{j}: |\sin\xi_{J}|> \frac32 \varepsilon_J^{\frac{1}{20}}\}$ has at most $2|\ln\varepsilon_0|^{2J^2 d}$ connected components, on which $\xi_{J}$ is smooth with $\xi_{J}'=-\frac{({\rm tr}A_{J})'}{2\sin\xi_{J}}$. If $J\geq 1$, then, on $\{\Gamma^{(J)}_{j+1}: |\sin\xi_{J}|> \frac32 \varepsilon_J^{\frac{1}{20}}\}$, $0\leq j\leq J-1$,
\begin{equation}\label{esti_plat}
\frac13< \xi_{J}' \leq \frac{N_j^{10\tau}}{|\sin\xi_{J}|} \ , \qquad \frac{\varepsilon_{j}^{\frac{3\sigma}{4}}}{4|\sin\xi_{J}|^3}< |\xi_{J}'' |\leq \frac{N_j^{20\tau}}{|\sin\xi_{J}|^{3}} \ .
\end{equation}

 \smallskip

\noindent
{\bf (S4)}
$\left|\rho(\{(\inf\Sigma,\sup\Sigma): |\sin\xi_{J}\right|\leq \frac32 \varepsilon_J^{\frac{1}{20}} \})|\leq \varepsilon_J^{\frac{1}{24}}$ and for $0\leq j\leq J$, $|\xi_{J}(\Gamma^{(J)}_j\setminus\Sigma_j)|\leq \varepsilon_{J}^{\frac{7\sigma}{8}}$.
\end{Theorem}

From now on, we denote $\rho_J:=\xi_{J} + \sum_{l=0}^{J-1}\lkr{ k_l}$,
which gives an approximation of $\rho$. In particular, $\rho_0=\xi_0$,
and
\begin{equation}
  \label{rhoj}
\left|\rho_{J}-\rho\right|_{\Sigma_j}\leq \varepsilon_j^{\frac14} \ .
\end{equation}

\noindent{\it Scheme of the proof of Theorems \ref{propsana} and
  \ref{propsana1}.}  The procedure of proof is a KAM procedure in
which one increases iteratively the order of the time dependent part
of the cocycle. The main point is that, in order to get a quite
complete description of the spectrum, one has to do the construction
for a set of $E$'s which is of full measure (not only of large
measure).  It is well known that the conjugacy of $A_0+F_0$ to a time
independent cocycle can be obtained through a close to identity
transformation only if some non-resonant relations are fulfilled
and this is typically true only in sets of large measure. To
describe the non-resonance condition, consider first the eigenvalues of
$A_0$: they can be written in the form $e^{\pm\im\rho_0}$, with
$\rho_0=\rho_0(E):=\arcos\left(-\frac{E}{2}\right)$, $|E|\leq 2$.  The
relevant non-resonance condition in order to construct the first
transformation is
\begin{equation}
\label{nonres0}
\left|\rho_0(E)-\lkr k\right|\geq
\frac{\varepsilon_0^\sigma}{\left|k\right|^\tau} \ ; \quad 0< |k|\leq N_0 \ .
\end{equation}
For the values of $E$ s.t. \eqref{nonres0} is fulfilled, the classical
construction of the KAM step produces a close to identity
transformation which conjugates $A_0+F_0$ to $A_1+F_1$ with
$|F_1|\sim\varepsilon_1$ and $A_1\in SL(2,\R)$ having eigenvalues of the
form $e^{\pm\im\rho_1} $, with $\rho_1$ close to $\rho_0$. Such a set
of $E$'s is $\Gamma^{(1)}_0$.

Now, let $k$ with $0<|k|\leq N_0$ be s.t. there exists a segment
$\cI_k$, on which equation \eqref{nonres0} is violated, then it is known
how to construct a time dependent matrix $H_{k,A_0}$ (which is not
close to identity) conjugating $A_0+F_0$ to a new cocycle $\tilde
A_0+\tilde F_0$, where $\tilde A_0$ has eigenvalues $e^{\pm\im \tilde \rho_0}$, with $\tilde\rho_0:=\rho_0-\lkr
k$. Furthermore, by the fact that $\omega$ is Diophantine, there are
no $\tilde k$ with $\tilde k\neq k$ s.t. \eqref{nonres0} is violated for $E\in\cI_k$. It
follows that \eqref{nonres0} is fulfilled by $\tilde\rho_0$ and
therefore, on $\cI_k$ one can conjugate $\tilde A_0+\tilde F_0$ to a
new cocycle $A_1+F_1$ with $|F_1|\sim\varepsilon_1$, and $A_1$ having
eigenvalues of the form $e^{\pm\im\alpha_1} $, with $\alpha_1$ close
to $\tilde\rho_0$, which in turn is close to $0$. It follows that for
some values of $E$, the quantity $\alpha_1$ can fail to be real. The
values of $E$ s.t. $\alpha_1(E)$ is purely imaginary are outside the
approximate spectrum of $H_\theta$, while the others belong to the approximate
spectrum. We put $\xi_1(E):=\Re(\alpha_1(E))$ and
$\rho_1(E):=\xi_1(E)+\lkr k$.

The union of the intervals $\cI_k$ is the set $\Gamma^{(1)}_1$.

In order to iterate we proceed as follows. For $E\in\Gamma^{(1)}_0$
one considers the non-resonance condition
\begin{equation}
\label{nonres1}
\left|\rho_1(E)-\lkr k\right|\geq
\frac{\varepsilon_1^\sigma}{\left|k\right|^\tau}\ ; \quad \forall \ 0< |k|\leq N_1\ .
\end{equation}
The set of the $E\in\Gamma^{(1)}_0$ for which (\ref{nonres1}) is satisfied is
$\Gamma^{(2)}_0$ and here one can construct a close to identity
transformation conjugating $A_1+F_1$ to $A_2+F_2$ with
$|F_2|\sim\varepsilon_2$ and $A_2\in SL(2,\R)$ having eigenvalues of the
form $e^{\pm\im\rho_2} $, with $\rho_2$ close to $\rho_1$.

Consider an element $E\in\Gamma^{(1)}_0$ s.t.  \eqref{nonres1} is
violated for some $k$. Such $E$'s are the first part of
$\Gamma^{(2)}_2$. For such $E$'s, one proceed as we did at the first
step in $\Gamma^{(1)}_1$.

Consider now $\Gamma^{(1)}_1$.
For these values of $E$ the relevant
non-resonance condition is
\begin{equation}
\label{nonres2}
\left|\xi_1(E)-\lkr k\right|\geq
\frac{\varepsilon_1^\sigma}{\left|k\right|^\tau}\ ; \quad \forall \  0< |k|\leq N_1\ .
\end{equation}
If it is fulfilled one proceeds as in $\Gamma^{(1)}_0$, namely, one
constructs a close to identity
transformation conjugating $A_1+F_1$ to $A_2+F_2$ with
$|F_2|\sim\varepsilon_2$ and $A_2\in SL(2,\R)$ having eigenvalues of the
form $e^{\pm\im\xi_2} $, with $\xi_2$ close to $\xi_1$. Such $E$'s
constitute $\Gamma^{(2)}_1$.

Consider now the $E\in\Gamma^{(1)}_1$ s.t.
$\exists \ k$ with $0<|k|\leq N_1$ s.t. \eqref{nonres2} is violated. The
union of such $E$'s is the remaining part of $\Gamma^{(2)}_2$. Here
one proceeds as we did for the first step in
$\Gamma^{(1)}_1$. Iterating and adding the estimates one gets the
proof of Theorem \ref{propsana1}.

In order to get Theorem \ref{propsana} one has simply to pass to the
limit $J\to \infty$. We do not discuss such a limit, which is
standard, but just recall that the sets $\Sigma_j$ are defined as
{\begin{equation}
  \label{sigmaj}
\Sigma_j:=\bigcap_{J:J\geq j}\Gamma^{(J)}_j\setminus \bigcup_{k\in\Z^d} \rho^{-1}(\lkr k) \ .
\end{equation}}

\subsection{Spectral
  transform}\label{spectrans}
\noindent

For $E\in\Sigma$, let $\cK(E)$ and $\cJ(E)$ be two linearly independent generalized
eigenvectors of $H_\theta$ and consider the spectral transform $\cS q$ defined
as follows: for any $q\in\ell^2(\Z)$, put
\begin{equation}
  \label{spec}
(\cS q)(E):=\left(\begin{array}{c}
                                                         \sum_{n}q_n \cK_n(E) \\[1mm]
                                                         \sum_{n}q_n \cJ_n(E)
                                                       \end{array}
     \right) \ .
\end{equation}
Given any matrix of measures on $\R$, namely $d\varphi=\left(\begin{array}{cc}
                 d\varphi_{11} & d\varphi_{12} \\[1mm]
                 d\varphi_{21} & d\varphi_{22}
               \end{array}\right)$,
let ${\cL}^2(d\varphi)$ be the space of the vectors
$G=(g_j)_{j=1,2}$, with $g_j$ functions of $E\in\R$ satisfying
\begin{equation}\label{gen_L2}
\|G\|_{{\cL}^2(d\varphi)}^2:=\sum_{j,k=1}^2 \int_\R g_j \, \bar g_k \,  d\varphi_{jk}<\infty \ .
\end{equation}

\begin{Theorem}[Chapter 9 of \cite{CL}]\label{spectral_measure_matrix}
There exists a Hermitian matrix of measures $\mu=(\mu_{jk})_{j,k=1,2}$, with $\mu_{jk}$ non-decreasing functions, such that $\cS:\ell^2(\Z)\to \cL^2(d\mu)$ is
unitary.
\end{Theorem}

Remark that by this theorem the spectral transform is invertible. As
anticipated in the introduction it is not known how to construct the
measure $d\mu$, however in \cite{Zhao} a procedure to construct an
approximate measure was developed.

Recall that $\sigma=\frac{1}{200}$, $|V|_r=\varepsilon_0< \varepsilon_{*}$ (as in Theorems \ref{propsana} and \ref{propsana1}) and the sequence $\{\varepsilon_{j}\}_j$ is defined by $\varepsilon_{j+1}=\varepsilon_{j}^{1+\sigma}$.

\begin{Proposition}
On the full measure subset
$\widetilde{\Sigma}:=\bigcup_{j\geq0}\Sigma_j$ of the spectrum, for any fixed $\theta\in \T^d$, any $E\in\widetilde{\Sigma}$, there exist two linearly independent
generalized eigenvectors ${\cK}(E)$ and ${\cJ}(E)$ of $H_\theta$
with the following properties: define the spectral transform according
to \eqref{spec} and consider the matrix of measures $d\varphi$ given by
$$
\left . d\varphi \right|_\Sigma :=\frac1\pi\left(\begin{array}{cc}
                                         \rho' & 0 \\[1mm]
                                         0 & \rho'
                                       \end{array}\right)\, dE  \ , \qquad \left. d\varphi\right|_{\R\setminus \Sigma} := 0 \ ,
$$
then we have,  for any $q\in\ell^2(\Z)$,
\begin{equation}\label{well-defined}
\left(1-\varepsilon_0^{\frac{\sigma^2}{10}}\right)\|q\|^2_{\ell^2(\Z)}\leq \|{{\cS}}q\|^2_{{\cL}^2(d\varphi)}\leq\left(1+\varepsilon_0^{\frac{\sigma^2}{10}}\right)\|q\|^2_{\ell^2(\Z)} \ ,
\end{equation}
and also
\begin{equation}
  \label{infinito}
\left| \frac{1}{\pi}\int_\Sigma \left(g_1(E){\cK}_n(E)+g_2(E){\cJ}_n(E)\right) \rho' dE -q_n\right|\leq
\varepsilon_0^{\frac{\sigma^2}{10}}\|q\|_{\ell^\infty} \ .
\end{equation}
Furthermore, the functions ${\cK}(E)$ and ${\cJ}(E)$ have the
following properties:
\begin{equation}\label{K_and_J}
{\cK}_n(E)=\sum_{n_{\Delta}=n,n\pm1}\beta_{n, n_{\Delta}}(E)\sin n_{\Delta}\rho(E) \ , \quad {\cJ}_n(E)=\sum_{n_{\Delta}=n,n\pm1}\beta_{n,n_{\Delta}}(E)\cos n_{\Delta}\rho(E) \ ,
\end{equation}
with $\rho$ the fibered rotation number of the cocycle $(\omega, A_0+F_0)$ and $$|\beta_{n, n_{\Delta}}-\delta_{n,n_{\Delta}}|_{\Sigma_0}\leq \varepsilon_0^{\frac14} \ , \qquad |\beta_{n, n_{\Delta}}|_{\Sigma_{j+1}}\leq \varepsilon_j^\sigma, \quad j\geq 0 \ .$$
Given any $J\in\N$, there exist $\beta^{J}_{n, n_{\Delta}}$, smooth on each connected component of $\Gamma^{(J)}_j$, satisfying
\begin{equation}\label{esti_beta_0}
\left|\beta^{J}_{n,n_\Delta} -\delta_{n, n_\Delta}\right|_{{\cC}^2(\Gamma^{(J)}_0)}\leq \varepsilon_{0}^{\frac14} \ ,
\end{equation}
and if $J\geq 1$, then
\begin{equation}\label{esti_beta_j+10}
|\beta_{n,n_\Delta}^{J}|_{{\cC}^1(\Gamma^{(J)}_{j+1})}\leq \varepsilon_j^{3\sigma} \ ,\quad 0\leq j\leq J-1 \ .
 \end{equation}
Moreover,
\begin{equation}\label{error3}
\left|\beta_{n, n_{\Delta}}-\beta^{J}_{n, n_{\Delta}}\right|_{\Sigma_j}\leq 10\varepsilon_{J}^{\frac14} \ ,\quad 0\leq j\leq J \ .
\end{equation}
\end{Proposition}

\noindent{\it Idea of the proof.} The construction and the estimates
of ${{\cS}}$ are actually given in Section 4.2 of \cite{Zhao}.  The
generalized eigenvectors ${\cK}$ and ${\cJ}$ are constructed as
Bloch waves exploiting the reducibility procedure and in particular
the matrices $Z$ and $B$ of Theorems \ref{propsana} and
\ref{propsana1}. The construction naturally leads to a family of
generalized eigenfunctions which do not depend in a smooth way on $E$
(in particular Eq. \eqref{esti_beta_0} and \eqref{esti_beta_j+10} do
not hold) so one modifies the normalization in order to get such
properties. The price to pay is that $\cS$ is no more unitary, but
turns out to be just a bounded transformation with bounded
inverse. For completeness we add now the details of the construction of
${\cK}$ and ${\cJ}$, while we refer to \cite{Zhao} for the details
of the proofs of the estimates.

{First we remark that one can construct Bloch-waves of Schr\"odinger
operator $H_\theta$ on $\widetilde{\Sigma}$ using the reducibility of Schr\"odinger cocycle.
Indeed, with an additional transform (see (3.17) of \cite{Zhao}), one can find $\tilde Z:\widetilde
\Sigma \times2\T^d\to SL(2,\C)$ and $B:\widetilde
\Sigma\to SL(2,\C)$, with two eigenvalues $e^{\pm{\rm i}\rho}$, such that
$$\tilde Z(\cdot+\omega)^{-1} (A_0+F_0(\cdot)) \, \tilde Z(\cdot)=\tilde B \ .$$
With the matrices
$\tilde Z=\left(\begin{array}{cc}
\tilde Z_{11} & \tilde Z_{12} \\
\tilde Z_{21} & \tilde Z_{22}
\end{array}\right)$ and
$\tilde B=\left(\begin{array}{cc}
\tilde B_{11} &\tilde B_{12} \\
\tilde B_{21} & \tilde B_{22}
\end{array}\right)$, one can easily see that,
defining}
\begin{align*}
\tilde
f_n(\theta)&:=\left[\tilde Z_{11}(\theta-\omega+n\omega)\tilde B_{12}-\tilde Z_{12}(\theta-\omega+n\omega)\tilde B_{11}\right]e^{-{\rm
    i}\rho}+\tilde Z_{12}(\theta-\omega+n\omega)
\\
\tilde\psi_n&=e^{{\rm i}n\rho}\tilde f_n(\theta) \ ,
\end{align*}
then such a $\tilde \psi$ fulfills $H_\theta\tilde\psi=E\tilde\psi$
for $E\in\widetilde\Sigma$. In order to get smooth dependence on $E$ we modify
its normalization in $\Sigma_j$ for $j\geq1$, defining
$$
\psi_n=e^{{\rm i}n\rho}f_n \;\ {\rm with} \;\ f_n=\left\{\begin{array}{ll}
 \tilde f_n \ , & E\in\Sigma_0 \\[1mm]
\tilde f_n\sin^5\xi, & E\in\Sigma_{j+1} \ ,  \;\  j\geq0
\end{array} \right. \ .
$$
Then we define ${\cK}_n:={\Im}(e^{{\rm i}n\rho}f_n \bar f_0)$ and
${\cJ}_n :={\Re}(e^{{\rm i}n\rho}f_n \bar f_0)$ on $\widetilde\Sigma$ and ${\cK}_n|_{\R\setminus \widetilde\Sigma}={\cJ}_n|_{\R\setminus\widetilde\Sigma}:=0$. By a direct calculation, we see
$$
e^{{\rm i}n\rho}f_n \bar f_0  =\sum_{n_\Delta= n, n\pm 1} \beta_{n,n_\Delta}e^{{\rm i}n_\Delta\rho} \ ,
$$ with some $\beta_{n,n_\Delta}$ which can be shown to fulfill the
estimates claimed in the statement (for the details see \cite{Zhao}).
Thus one gets ${\cK}_n=\sum_{n_\Delta}\beta_{n, n_\Delta} \sin
n_\Delta\rho$, ${\cJ}_n=\sum_{n_\Delta}\beta_{n, n_\Delta} \cos
n_\Delta\rho$.

Finally one has to show the important estimates \eqref{well-defined}
and \eqref{infinito}. They
were proved in \cite{Zhao}. Here we just recall that the main step for
its proof are the following inequalities
\begin{eqnarray*}
\left|\frac{1}{\pi}\int_\Sigma \left({\cK}^2_n(E)+{\cJ}^2_n(E)\right)  \, \rho'  dE -1\right|&\leq& \varepsilon_0^{\frac{\sigma^2}{8}} \ ,\\
\left|\frac{1}{\pi}\int_\Sigma \left({\cK}_m(E){\cK}_n(E)+{\cJ}_m(E){\cJ}_n(E)\right) \, \rho' dE \right|&\leq& \frac{\varepsilon_0^{\frac{\sigma^2}{8}} }{|m-n|^{1+\frac{\sigma}{6}}} \ , \quad m\neq n \ ,
\end{eqnarray*}
the second of which is obtained from an estimate of an oscillatory
integral (Lemma 4.1 of \cite{Zhao}) which is very close to the
estimate given in Lemma \ref{lemma_Mlarge} of the present paper.\qed

\section{An oscillatory integral on the spectrum}\label{sec_int}
\noindent

In this section, by using the division
of $[\inf\Sigma,\sup\Sigma]$ given in Theorem \ref{propsana1}, we estimate an integral on the spectrum.
This will be applied in analyzing the time evolution, and deducing dispersion in the next section.

Recall that $\sigma=\frac{1}{200}$, $|V|_r=\varepsilon_0\leq \varepsilon_{*}$ and the sequence $\{\varepsilon_{j}\}_{j\geq0}$ is defined by $\varepsilon_{j+1}=\varepsilon_{j}^{1+\sigma}$.

\begin{Lemma}\label{h_integral_t}
Let $h:\widetilde\Sigma\to\R$ be a function s.t. for any
$J\geq 0$ there exists a function $h_J:\bigcup_{0\leq j\leq
  J}\Gamma^{(J)}_j\to \R$ which is ${\cC}^1$ on each connected
component of $\Gamma^{(J)}_j$, and satisfies the following assumptions
\begin{itemize}
\item[(E0)] $\left|h_J-h\right|_{\Sigma_j}\leq 10\varepsilon_J^{\frac14}$ for $0\leq j \leq J$ and $\left|h_J-h\right|_{\Sigma_j}\leq 2$ for $j \geq J+1$,\\[1mm]
\item [(E1)] $\displaystyle |h_J|_{{\cC}^1(\Gamma^{(J)}_0)}\leq \frac{16}{15}$,\\[1mm]
  \item [(E2)] $\displaystyle |h_J|_{{\cC}^1(\Gamma^{(J)}_{j+1})} \leq \varepsilon_j^{3\sigma}$, $0\leq j\leq J-1$, if $J\geq 1$.
\end{itemize}
Then, there exists a positive constant $a<80802$ s.t. for any
$M\in\R$, one has
\begin{equation}
  \label{esti_osc}
\left|\int_{\Sigma} h e^{-{\rm i}E t}\, \cos M \rho\cdot    \rho'\,
dE\right|\leq\frac{ 526 |\ln\varepsilon_0|^{a(\ln\ln(2+\langle t\rangle))^2 d}} {\langle t\rangle^{\frac13}} \ .
\end{equation}
\end{Lemma}

The rest of the section is devoted to the proof of such a lemma.

From now on we assume that $\varepsilon_*$ is such that all the smallness
conditions that we will assume are satisfied.

We denote
$${\cI}_M({\cS}):=\int_{{\cS}} h e^{-{\rm i}E t}\, \cos M \rho\cdot    \rho'\, dE, \quad {\cS}\subset\R \ ,$$
and
$${\cI}_M^{J}({\cS}):=\int_{{\cS}} h_J e^{-{\rm i}E t}\, \cos
M \rho\cdot \rho'\, dE, \quad {\cS}\subset\R \ .$$
We first give three
lemmas, the first of which allows to approximate $\cI_M$ through
$\cI_M^J$. For the estimate of $\cI_M^J$, we have to separate the
cases of small $M$ and large $M$: they are treated in two different
lemmas. Finally we will summarize the results and deduce Lemma
\ref{h_integral_t}.

\begin{Lemma}
  \label{approx}
For any positive $J$ and any $M\in\R$, under the assumptions of Lemma
\ref{h_integral_t}, one has
  \begin{equation}
    \label{approx.1}
\left|\cI_M(\Sigma)-\cI_M^J(\Sigma)\right|\leq
\varepsilon_J^{\frac{3\sigma}{4}} \ .
  \end{equation}
\end{Lemma}
\proof By the fact that
$|\rho(\Sigma_{j+1})|\leq 3|\ln\varepsilon_j|^{2
  d}\varepsilon_j^{\sigma}$, we have
\begin{eqnarray*}
\sum_{j=0}^{J}\left|\int_{\Sigma_j}(h-h_J )
\cos (M\rho) \cdot e^{-{\rm i}Et} \rho' \,
dE\right|&\leq&10\varepsilon_J^{\frac14}\sum_{j=0}^{J}\int_{\Sigma_j} \left|\rho' \right|
dE\\
&=&10\varepsilon_J^{\frac14}\sum_{j=0}^{J}\int_{\Sigma_j} \rho'
dE
\\
&\leq& 10\varepsilon_J^{\frac14}\sum_{j=0}^{J} \left|\rho(\Sigma_j)\right|\\
&\leq& 10\varepsilon_J^{\frac14}\left(2\pi+\sum_{j\geq0}3\varepsilon_j^{\sigma}\left|\ln
\varepsilon_j\right|^{2d}\right)\\
&\leq&10\varepsilon_J^{\frac14}(2\pi+4\varepsilon_0^{\sigma})\\
&\leq&\frac12\varepsilon_J^{\frac16}
\end{eqnarray*}
and $$\sum_{j\geq J+1}\left|\int_{\Sigma_j}(h-h_J ) \cos(M\rho) \cdot  e^{-{\rm i}Et}  \rho' \,  dE\right|\leq \frac12\varepsilon_J^{\frac{3\sigma}{4}} \ .$$
Hence we get that the error is bounded by
\begin{equation}\label{int_appro}
 \frac12\varepsilon_J^{\frac16}+\frac12\varepsilon_J^{\frac{3\sigma}{4}}\leq \varepsilon_{J}^{\frac{3\sigma}{4}} \ . \qed
\end{equation}

\begin{Lemma}\label{lemma_Mlarge}
Assume that for some positive $J\geq0$ the function $h_J$ fulfills (E2)
and (E3), then for every $M\in\R\setminus\{0\}$ and $t\in\R$, we have
\begin{equation}
  \label{mlarge.1}
\left|\cI_M^J(\Sigma)\right|\leq \frac{32}{15}\frac{1}{\left|M\right|}
\left|\ln
  \varepsilon_0\right|^{2J^2d}+\frac{32}{15}\frac{1}{\left|M\right|} (\sup\Sigma-\inf\Sigma)\langle
  t\rangle \ .
\end{equation}
\end{Lemma}
\proof Since $\rho'=0$ on $[\inf\Sigma,\sup\Sigma]\setminus\Sigma$, then we have
$${\cI}_M^J(\Sigma)=\int_{\inf\Sigma}^{\sup\Sigma} h_Je^{-{\rm i}E t}\, \cos M \rho   \cdot \rho'\, dE \ .$$
The above integral on the right hand side is indeed the sum of integrals over the connected component $(E_*,\, E_{**})\subset \Gamma^{(J)}_j$.
Since $\rho$ is absolutely continuous, by integrating by parts on each
connected component, we obtain
\begin{eqnarray*}
& &\int_{\inf\Sigma}^{\sup\Sigma} h_J e^{-{\rm i}Et}\,  \cos M \rho   \cdot \rho' \, dE\\
&=& \frac{1}{M}\sum_{j=0}^{J}\sum_{(E_*, \, E_{**})\subset \Gamma^{(J)}_j \atop{\rm connected \ component}}\left. h_J e^{-{\rm i}Et} \, \sin M \rho\right|_{(E_*, E_{**})} \\
& &- \,  \frac{1}{M}\sum_{j=0}^{J}\sum_{(E_*, \, E_{**})\subset \Gamma^{(J)}_j \atop{\rm connected \ component}}\int_{E_*}^{E_{**}} (h_Je^{-{\rm i}Et})'\sin M \rho\,  dE \ .
\end{eqnarray*}
Since there are at most $|\ln\varepsilon_0|^{2J^2 d}$ connected components of $\bigcup_{j=0}^J\Gamma^{(J)}_j$, we have
\begin{eqnarray*}
\frac{1}{|M|}\left|\sum_{j=0}^{J}\sum_{(E_*, \, E_{**})\subset
  \Gamma^(J)_j \atop{\rm connected \, component}}\left. h_J e^{-{\rm
    i}Et} \, \sin M \rho\right|_{(E_*, E_{**})}\right|\leq
\frac{32}{15 |M|}|\ln\varepsilon_0|^{2J^2d}
\end{eqnarray*}
and
\begin{eqnarray*}
\frac{1}{|M|}\left|\sum_{j=0}^{J}\sum_{(E_*, \, E_{**})\subset \Gamma^{(J)}_j \atop{\rm connected \, component}}\int_{E_*}^{E_{**}} (h_Je^{-{\rm i}Et})'\sin M \rho\,  dE\right|
\leq\frac{32|t|}{15 |M|}( \sup\Sigma-\inf\Sigma) \ . \qed
\end{eqnarray*}

\begin{Lemma}
\label{lemma_Msmall}
Assume that for some positive $J\geq0$ the function $h_J$ fulfills (E2)
and (E3), then for every $M\in\R$ and $t\in\R$, we have
\begin{equation}\label{msmall.1}
\left|\cI_M^J(\Sigma)\right|\leq 512
\frac{\left|\ln\varepsilon_0\right|^{2J^2d}}{\langle t\rangle^{\frac13}}+\frac{1}{2}\varepsilon_J^{\frac{3\sigma}{4}}+2\left|M\right| \varepsilon_J^{\frac{1}{4}} \ .
\end{equation}
\end{Lemma}
\proof The proof is divided into three parts.

\smallskip

{\bf Step 1. Approximation}

\smallskip

We will consider the sum of integrals
$$\sum_{j=0}^{J} \int_{\left\{\Gamma^{(J)}_j:{|\sin\xi|>\varepsilon_J^{\frac{1}{20}}}\right\}} h_J e^{-{\rm i}E t}\, \cos M \rho_{J}   \cdot \rho_{J}'\, dE$$
instead of ${\cI}_M^J(\Sigma)$. The error is estimated by
\begin{eqnarray}
& & \left| {\cI}_M^J(\Sigma)-\sum_{j=0}^{J}\int_{\left\{\Gamma^{(J)}_j:{|\sin\xi|>\varepsilon_J^{\frac{1}{20}}}\right\}}
  h_J e^{-{\rm i}E t}\, \cos M \rho_{J} \cdot \rho_{J}'\, dE
  \right|\label{error_int}\\ &\leq&\left|\sum_{j=0}^{J}
  \int_{\left\{\Sigma_{j}:|\sin\xi|>
    \varepsilon_J^{\frac1{20}}\right\}} h_Je^{-{\rm i} E t}\left(\cos
  M\rho_{J}\cdot\rho_{J}' -\cos M\rho\cdot\rho' \right) \,
  dE\right|\label{error_int_01}\\ & & + \,
  \left|\sum_{j=0}^{J}\int_{\left\{\Gamma^{(J)}_j\setminus\Sigma_j:{|\sin\xi|>\varepsilon_J^{\frac{1}{20}}}\right\}}h_J
  e^{-{\rm i}E t} \, \cos M \rho_{J} \cdot \rho_{J}'\,
  dE\right| \label{error_int_02} \\ & & + \,
  \left|\sum_{j=0}^{J}\int_{\left\{\Sigma_{j}:|\sin\xi|\leq\varepsilon_J^{\frac{1}{20}}\right\}}h_J
  e^{-{\rm i}E t} \cos M\rho\cdot\rho' \,
  dE\right|\label{error_int_03}\\ & & + \, \left| \sum_{j\geq
    J+1}{\cI}_M^J(\Sigma_{j})\right| \ .  \label{error_int_04}
\end{eqnarray}
\begin{itemize}
\item Since $|\rho(\Sigma_{j+1})|\leq 3|\ln\varepsilon_j|^{2 d}\varepsilon_j^{\sigma}$,
the term in (\ref{error_int_04}) is bounded by
$$
\frac{32}{5}|\ln\varepsilon_J|^{2 d}\varepsilon_{J}^{\sigma} \leq \frac14 \varepsilon_J^{\frac{3\sigma}{4}}.
$$
\item On $\Sigma_{j}$, $0\leq j \leq J$, we have $|\xi_{J}-\xi|\leq \varepsilon_{J}^{
\frac14}$. So $|\sin\xi|\leq\varepsilon_J^{\frac{1}{20}}$ implies that $|\sin\xi_{J}|\leq\frac32\varepsilon_J^{\frac{1}{20}}$.
By the assertion {\bf (S4)} of Theorem \ref{propsana1}, the term in (\ref{error_int_03}) is bounded by
$$\frac{16}{15} \varepsilon_{J}^{\frac{1}{24}}\leq \varepsilon_{J}^{5\sigma}.$$
\item By the fact that $|\rho_{J}(\Gamma^{(J)}_j\setminus\Sigma_j)|\leq \varepsilon_{J}^{\frac{7\sigma}8}$, $0\leq j\leq J$, the term in (\ref{error_int_02}) is bounded by
$$\frac{16}{15}(J+1)\cdot \varepsilon_{J}^{\frac{7\sigma}8}\leq \frac14 \varepsilon_{J}^{\frac{3\sigma}4}.$$
\item On $\left\{\Sigma_{j}:|\sin\xi|> \varepsilon_J^{\frac1{20}}\right\}$, $0\leq j\leq J$, we have $|\xi_{J}-\xi|\leq \varepsilon_J^{\frac14}$, which implies $|\sin\xi_{J}|\geq \frac12 \varepsilon_{J}^{\frac{1}{20}}$. Then, by (\ref{sigma_m_0})--(\ref{error_whitney}), we get
\begin{eqnarray*}
|\rho_{J}'-\rho'|&=&\frac12\left|\frac{({\rm tr} A_{J})'}{\sin \xi_{J}}-\frac{({\rm tr} B)'}{\sin \xi}\right|\\
&=&\frac{|({\rm tr} A_{J})'\sin\xi-({\rm tr} B)'\sin\xi_{J}|}{2|\sin\xi||\sin\xi_{J+1}|}\\
&\leq&\frac{|({\rm tr} A_{J})'||\sin\xi-\sin\xi_{J}|+|\sin\xi_{J}||({\rm tr} B)'-({\rm tr} A_{J})'|}{2|\sin\xi||\sin\xi_{J}|}\\
&\leq& 2\varepsilon_{J}^{-\frac{1}{10}}\cdot \left(2 \varepsilon_J^{\frac14}N_J^{10\tau}+2\varepsilon_J^{\frac14}\right) \\
&\leq&\varepsilon_J^{\frac1{10}} \ ,
\end{eqnarray*}
and, using
$$|\cos M\rho_{J}-\cos M\rho|=2\left|\sin\frac{M}{2}(\rho_{J}+\rho)\right|\left|\sin\frac{M}{2}(\rho_{J}-\rho)\right|
\leq2 |M|\cdot\varepsilon_J^{\frac14}
$$
we bound of the term (\ref{error_int_01}).
\end{itemize}
Hence, by combining the above estimates, the error given in
(\ref{error_int}) is less than
$$\frac12\varepsilon_J^{\frac{3\sigma}4}+2\left|M\right|\varepsilon_J^{\frac{1}{4}} \ .$$

{\bf Step 2. Change of variable}\\

Recall that there are at most $2|\ln\varepsilon_0|^{2J^2d}$ connected  components of
$$\bigcup_{j=0}^J\left\{E\in\Gamma^{(J)}_j:|\sin\xi_{J}|>\frac32\varepsilon^{\frac1{20}}_J\right\} \ .$$
Let $(E_*, E_{**})$ be one of these components, on which $\rho_{J}(E)$ is strictly increasing.
So $E=E(\rho_{J})$ is well-defined with
\begin{equation}\label{derivative_inverse}
\frac{d E}{d\rho_{J}}=\frac{1}{\rho'_{J}} \ ,\quad \frac{d^2 E}{d\rho_{J}^2}=-\frac{\rho''_{J}}{(\rho'_{J})^3} \ ,\quad \frac{d^3 E}{d\rho_{J}^3}=\frac{3(\rho''_{J})^2}{(\rho'_{J})^5}-\frac{\rho'''_{J}}{(\rho'_{J})^4} \ .
\end{equation}
Since $\rho_{J}'=\xi_J'> \frac13$, we have $|\frac{d E}{d\rho_{J}}|<3$.
Then, for $F(\rho_{J}):=(h_J\circ E)(\rho_{J})$, in view of the condition (E2) and (E3) for $h_J$, we can get
\begin{equation}\label{esti_F}
\left| F\right|_{{\cC}^1(\rho_{J}(\Gamma^{(J)}_0))}\leq \frac{16}{5} \ ; \qquad |F|_{{\cC}^1(\rho_{J}(\Gamma^{(J)}_{j+1}))} \leq 3\varepsilon_j^{3\sigma} \ ,\quad 0\leq j\leq J-1 \ ({\rm if} \ J\geq 1 ) \ .
\end{equation}
By the change of variable, we get
\begin{eqnarray}
& &\int_{E_*}^{E_{**}} h_Je^{-{\rm i}E t} \, \cos M \rho_{J}   \cdot \rho_{J}'\, dE\nonumber\\
&=&\frac12\int_{\rho_{J}(E_*)}^{\rho_{J}(E_{**})} F(\rho_{J}) \left(e^{-{\rm i} t\left[E(\rho_{J})+\frac{M}{t} \rho_{J}\right]}+e^{-{\rm i} t\left[E(\rho_{J})-\frac{M}{t} \rho_{J}\right]}\right)\, d\rho_{J} \ .\label{int_seg_gamma}
\end{eqnarray}

{\bf Step 3. Van der Corput lemma on each component}\\

We first prove that, for any $E\in(E_*, E_{**})\subset\Gamma^{(J)}_0$,
we have
\begin{equation}
  \label{3.2}
\text{either}\quad \left|\frac{d^2 E}{d\rho_{J}^2}\right|\quad\text{
  or}\quad
\left|\frac{d^3 E}{d\rho_{J}^3}\right|\geq
  1-\varepsilon_0^{\frac13}\ .
\end{equation}

By (\ref{sigma_m_0}) and the fact that
$A_0=\left(
\begin{matrix}
  -E & -1 \\
   1 & 0
\end{matrix}\right)
$, we can see
$$|({\rm tr} A_{J})'+1| \ , \; |({\rm tr} A_{J})''| \ , \; |({\rm tr} A_{J})'''|\leq 2\varepsilon_0^{\frac12} \ {\rm on} \ \Gamma_0 \ .$$
Since $\rho_{J}=\xi_{J}$ on $\Gamma^{(J)}_0$ and $\xi_{J}'=-\frac{({\rm tr} A_{J})'}{2\sin\xi_{J}}$,
combining with (\ref{derivative_inverse}), we have
\begin{eqnarray*}
\frac{d^2 E}{d\rho_{J}^2}&=&-\frac{4({\rm tr} A_{J})''\sin^2\rho_{J}}{({\rm tr} A_{J})'^3}-\frac{2\cos\rho_{J}}{({\rm tr} A_{J})'}\\
\frac{d^3 E}{d\rho_{J}^3}
&=&-\frac{24({\rm tr} A_{J})''^2\sin^3\rho_{J}}{({\rm tr} A_{J})'^5}+\frac{8({\rm tr} A_{J})'''\sin^3\rho_{J}}{({\rm tr} A_{J})'^4} -\frac{12({\rm tr} A_{J})''\cos\rho_{J}\sin\rho_{J}}{({\rm tr} A_{J})'^3}+\frac{2\sin\rho_{J}}{({\rm tr} A_{J})'} \ ,
\end{eqnarray*}
and hence
\begin{equation}\label{transversality0}
\left|\frac{d^2 E}{d\rho_{J}^2}\right|+\left|\frac{d^3 E}{d\rho_{J}^3}\right|\geq\frac{2}{|({\rm tr} A_{J})'|}\left(|\cos\rho_{J}|+|\sin\rho_{J}|\right)- 40\varepsilon_0^{\frac12}\nonumber\\
\geq 2(1-\varepsilon_0^{\frac13}) \ ,
\end{equation}
which implies \eqref{3.2}.

Let ${\cJ}\subset (E_*,E_{**})$ be the subset such that
$$\left|\frac{d^3 E}{d\rho_{J}^3}\right|\geq 1-\varepsilon_0^{\frac13} \ {\rm on} \ {\cJ} \ ; \quad
\left|\frac{d^3 E}{d\rho_{J}^3}\right|< 1-\varepsilon_0^{\frac13} \ {\rm on} \ (E_*,E_{**})\setminus {\cJ} \ .
$$
Since
\begin{eqnarray*}
& & \left|\frac{d^3 E}{d\rho_{J}^3}-2\sin\rho_{J}\right|\\
&\leq &|\sin\rho_{J}|\left[ 2\left(\frac{1}{1-2\varepsilon_0^{\frac12}}-1\right) + \frac{96\varepsilon_0}{(1-2\varepsilon_0^{\frac12})^5}  + \frac{16\varepsilon_0^{\frac12}}{(1-2\varepsilon_0^{\frac12})^4} +\frac{24\varepsilon_0^{\frac12}}{(1-2\varepsilon_0^{\frac12})^3}\right]\\
&\leq &\varepsilon_0^{\frac13} |\sin\rho_{J}| \ ,
\end{eqnarray*}
and $\rho_{J}\in
\left(-\varepsilon^{\frac14}_J,\pi+\varepsilon^{\frac14}_J\right)$,
one has that ${\cJ}\subset (E_*,E_{**})$ is composed at most by one
sub interval (maybe empty) and $(E_*,E_{**})\setminus {\cJ}$
consists of at most two sub-intervals, saying ${\cS}_1$ and ${\cS}_2$.

On ${\cJ}$, we apply Van der Corput lemma (Corollary
\ref{coro_VDC}) with $k=3$, and get (for $|t|\geq 1$)
$$\left| \int_{\rho_{J}({\cJ})} F(\rho_{J}) e^{-{\rm i} t\left[E(\rho_{J})\pm\frac{M}{t} \rho_{J}\right]} \, d\rho_{J}\right|\leq 18(1-\varepsilon_0^{\frac13})^{-\frac{1}{3}} \cdot \frac{16}{5}(1+\pi)
|t|^{-\frac13}\leq 240 |t| ^{-\frac13} \ .$$
On ${\cS}_1$ and ${\cS}_2$, we have $|\frac{d^2
  E}{d\rho_{J}^2}|\geq 1-\varepsilon_0^{\frac13}$ in view of
(\ref{transversality0}), then, by applying Corollary \ref{coro_VDC}
with $k=2$, we get, for $l=1,2$, and $|t|\geq 1$
$$\left| \int_{\rho_{J}({\cS}_l)} F(\rho_{J}) e^{-{\rm i} t\left[E(\rho_{J})\pm\frac{M}{t} \rho_{J}\right]} \, d\rho_{J}\right|\leq  8 (1-\varepsilon_0^{\frac13})^{-\frac{1}{2}}\cdot
 \frac{16}{5}(1+\pi) |t|^{-\frac12}\leq 108 |t| ^{-\frac12} \ .$$
Hence, the integral in (\ref{int_seg_gamma}) is bounded by $456 |t|^{-\frac13}$ for every connected component $(E_*,E_{**})$ contained in $\Gamma_{0}$.
Recalling that there are $|\ln\varepsilon_0|^{2J^2 d}$ connected components in $\Gamma_0$, we get
{\begin{eqnarray}\nonumber
\left|\int_{\left\{\Gamma_{0}:|\sin\xi|>\varepsilon_J^{\frac{1}{20}}\right\}}
h_Je^{-{\rm i}E t}\, \cos M \rho_{J} \cdot \rho_{J}'\, dE\right|
&\leq& |\ln\varepsilon_0|^{2J^2 d} \cdot 456 \, |t|^{-\frac13} \\ & \leq&
2^{\frac16}\cdot456 |\ln\varepsilon_0|^{2J^2d} \langle
t\rangle^{-\frac13} \ , \label{esti_0}
\end{eqnarray}
since $\frac{\sqrt{1+t^2}}{|t|}\leq 2^{\frac12}$ for $|t|\geq 1$. }

\smallskip

If $J\geq 1$, then for $(E_*, E_{**})\subset\Gamma^{(J)}_{j+1}$, $0\leq j\leq J-1$,
(\ref{esti_plat}) implies that
$$|\rho'_{J}|=|\xi'_{J}|\leq \frac{N_j^{10\tau}}{|\sin\xi_{J}|} \ , \quad |\rho_{J}''|=|\xi_{J}''|\geq \frac{\varepsilon_j^{\frac{3\sigma}{4}}}{4|\sin\xi_{J}|^3} \ ,$$
Hence the second derivative of the inverse function satisfies
\begin{equation}\label{transversality_j+1}
\left|\frac{d^2 E}{d\rho_{J}^2}\right| =\frac{|\rho_J''|}{|\rho_J'|^3}\geq
\frac{\varepsilon_j^{\frac{3\sigma}{4}}}{4|\sin\xi_{J}|^3}\cdot\frac{|\sin\xi_{J}|^3}{N_j^{60\tau}}> \varepsilon_{j}^{\frac{7\sigma}{8}} \ .
\end{equation}
So we apply Corollary \ref{coro_VDC} with $k=2$, and get
$$\left| \int_{\rho(E_*)}^{\rho(E_{**})} F(\rho_{J})  e^{-{\rm i} t\left[E(\rho_{J})\pm\frac{M}{t} \rho_{J}\right]}\, d\rho_{J}\right|\leq  8\varepsilon_{j}^{-\frac{7\sigma}{16}}\cdot
3\varepsilon_j^{3\sigma}(1+\pi)|t|^{-\frac12}\leq \varepsilon_j^{\frac{5\sigma}2} |t|^{-\frac12} \ .$$
Hence, we have
\begin{eqnarray}
\left|\sum_{j=0}^{J-1}\int_{\left\{\Gamma^{(J)}_{j+1}:|\sin\xi|>\varepsilon_J^{\frac{1}{20}}\right\}}
h_Je^{-{\rm i}E t} \, \cos M \rho_{J} \cdot \rho_{J}'\, dE\right|&\leq& |\ln\varepsilon_0|^{2J^2 d} \varepsilon_0^{\frac{5\sigma}2} |t|^{-\frac12} \nonumber \\
&\leq&
|\ln\varepsilon_0|^{2J^2 d} \varepsilon_0^{2\sigma} \langle
t\rangle^{-\frac13}\ . \label{esti_j+1}
\end{eqnarray}

\smallskip

By combining (\ref{esti_0}) and (\ref{esti_j+1}), we get, for $|t|\geq 1$,
{\begin{eqnarray*}
\left|\sum_{j=0}^J
  \int_{\left\{\Gamma^{(J)}_j:{|\sin\xi|>\varepsilon_J^{\frac{1}{20}}}\right\}}
  h_J e^{-{\rm i}E t}\, \cos M \rho_{J} \cdot \rho_{J}'\, dE\right|  &\leq&\left(2^{\frac16}\cdot 456+\varepsilon_0^{2\sigma}\right)|\ln\varepsilon_0|^{2J^2 d} \langle t\rangle^{-\frac13} \\
  &\leq& 512 |\ln\varepsilon_0|^{2J^2 d} \langle t\rangle^{-\frac13} \ .
\end{eqnarray*}
Since the above inequality holds trivially for $|t|\leq 1$,
this concludes the proof of Lemma \ref{lemma_Msmall}.}\qed

\smallskip

We are now ready for the

\noindent {\it Proof of Lemma \ref{h_integral_t}. } Fix $t$, and
choose $J$ in such a way that the error in Lemma \ref{approx}
satisfies
$$
\varepsilon_J^{\frac{3\sigma}{4}}\leq \frac{1}{\langle t\rangle^{\frac13}}\ .
$$
this gives
\begin{equation}
  \label{J}
J\geq
J_*:=\frac{1}{\ln(1+\sigma)}\ln\left(\frac{4}{9\sigma}\frac{\ln\langle
  t\rangle}{\left|\ln\varepsilon_0\right|}  \right) \ .
\end{equation}
Taking $J_{\sharp}$ to be the smallest integer fulfilling \eqref{J},
one has that, provided $\varepsilon_0$ is small enough, one has
{\begin{equation} \label{Jsh}
J_{\sharp}\leq J_*+1<\frac{1}{\ln(1+\sigma)}\ln\ln(2+\langle t\rangle)\leq
201 \ln\ln(2+\langle t\rangle) \ .
\end{equation}}

If $|M|\geq \frac{32}{5}\langle t\rangle^{\frac43}$, then we use the
estimate \eqref{mlarge.1}. In such a case, the second term at r.h.s of
\eqref{mlarge.1} is estimated by {$\frac{5}{3}\frac{1}{\langle
  t\rangle^{\frac13}}$}. The first term (with $J=J_\sharp$) is estimated by
{$\frac{\left|\ln\varepsilon_0\right|^{2J_\sharp^{2}d}}{3 \langle
  t\rangle^{\frac43}}$}. Summing up we get the result for the considered
values of $M$.

Consider now $|M|<\frac{32}{5}\langle t\rangle^{\frac43}$ and use
\eqref{msmall.1}. The first two terms at r.h.s. are immediately
estimated. For the third one just remark that
$$
2\left|M\right|\varepsilon_J^{\frac{1}{4}}\leq
\frac{64}{5}\langle t\rangle^{\frac43} \left(\varepsilon_J^{\frac{3\sigma}{4}} \right)^{\frac{1}{3\sigma}}
 \leq \frac{64}{5} \langle
t\rangle^{\frac43-\frac{1}{9\sigma}}=\frac{64}{5}\frac{1}{\langle
  t\rangle^{\frac{200}{9}-\frac{1}{3} }}
\leq \frac{64}{5}\frac{1}{\langle
  t\rangle^{20}} \ .
$$
Summing up one gets the result.\qed

\section{Proof of dispersive estimates}\label{sec_proof}

Fix any $\theta\in\T^d$.
Given $\phi\in\ell^1(\Z)$, let $q(t)=e^{-{\rm i}t H_{\theta}}\phi$. It solves the dynamical equation ${\rm i}\dot{q}=H_\theta q$ with $q(0)=\phi$.
Let $$G(E,t)\equiv\left(\begin{matrix}
g_1(E,t)\\ g_2(E,t)
\end{matrix}\right):={\cS}(q(t)).$$
For a.e. $E\in\Sigma$, we have
$\left(\begin{array}{c}
g_1(E,t) \\[1mm]
g_2(E,t)
\end{array}\right)= e^{-{\rm i}Et} \left(\begin{array}{c}
g_1(E,0) \\[1mm]
g_2(E,0)
\end{array}\right)$.

In view of eq. \eqref{infinito}, we have
\begin{equation}\label{q(t)_int}
|q_n(t)|\leq\frac{1}{\pi}\left|\int_\Sigma \left(g_1(E,t){\cK}_n(E)+g_2(E,t){\cJ}_n(E)\right) \rho' dE\right|+\varepsilon_0^{\frac{\sigma^2}{10}}\|q(t)\|_{\ell^\infty} \ ,\quad \forall \  n\in\Z \ .
\end{equation}
To estimate $\|q(t)\|_{\ell^\infty}$, it is sufficient to control the above integral.
By a straightforward computation, we have
\begin{eqnarray}
& &\int_\Sigma \left(g_1(E,t){\cK}_n(E)+g_2(E,t){\cJ}_n(E)\right)\, \rho'  dE \nonumber \\
&=&\int_\Sigma e^{-{\rm i}Et}\left(g_1(E,0){\cK}_n(E)+g_2(E,0){\cJ}_n(E)\right)\, \rho'  dE \nonumber \\
&=& \int_\Sigma e^{-{\rm i}Et}  \sum_{m\in\Z}\phi_m\left({\cK}_m(E){\cK}_n(E)+{\cJ}_m(E){\cJ}_n(E)\right)\, \rho'  dE \nonumber \\
&=& \int_\Sigma e^{-{\rm i}Et}  \sum_{m\in\Z} \phi_m \sum_{m_{\Delta}, n_{\Delta}}
\left(\beta_{m,m_{\Delta}} \beta_{n,n_{\Delta}} \cos(m_{\Delta}-n_{\Delta})\rho \right)\, \rho'  dE \ .\label{appro_q}
\end{eqnarray}
\begin{Lemma}\label{Lemma_int_appro}
Assume that $|V|_r=\varepsilon_{0}\leq \varepsilon_{*}$ with
$\varepsilon_{*}$ in Theorem \ref{propsana}.  For any
$m,m_{\Delta},n,n_{\Delta}$,
{$$\left|\int_\Sigma  \beta_{m,m_{\Delta}} \beta_{n,n_{\Delta}}
\cos(m_{\Delta}-n_{\Delta})\rho \cdot  e^{-{\rm i}Et}   \, \rho'  dE
\right|\leq \frac{526 |\ln\varepsilon_0|^{a(\ln\ln(2+\langle t\rangle))^2 d}} {\langle t\rangle^{\frac13}} \ , \quad \forall \ t\in\R \ .$$}
\end{Lemma}
\proof We just apply Lemma \ref{h_integral_t} with $h=
\beta_{m,m_{\Delta}} \beta_{n,n_{\Delta}}$, $M=m_{\Delta}-n_{\Delta} $
and $h_J=\beta^J_{m,m_{\Delta}} \beta^J_{n,n_{\Delta}}$. The result
immediately follows. \qed

\

\noindent{\it End of the proof of Theorem \ref{dispersion_log}.}
According to (\ref{appro_q}) and Lemma \ref{Lemma_int_appro}, we get, for every $n\in\Z$,
{ $$
\left|\int_\Sigma \left(g_1(E,t){\cK}_n(E)+g_2(E,t){\cJ}_n(E)\right)\, \rho'  dE\right|\leq \frac{9\cdot 526
    |\ln\varepsilon_0|^{a(\ln\ln(2+\langle t\rangle))^2 d}} {\langle
    t\rangle^{\frac13}} \|q(0)\|_{\ell^1} \ .
$$}
Finally, by (\ref{q(t)_int}), we get,  for every $t\in\R$,
\begin{eqnarray*}
\|q(t)\|_{\ell^\infty} &\leq&  \frac{9\cdot 526}{\pi \left(1-\varepsilon_0^{\frac{\sigma^2}{10}} \right) }\frac{
    |\ln\varepsilon_0|^{a(\ln\ln(2+\langle t\rangle))^2 d}} {\langle
    t\rangle^{\frac13}} \|q(0)\|_{\ell^1} \\
 &\leq&  1507  \frac{
    |\ln\varepsilon_0|^{a(\ln\ln(2+\langle t\rangle))^2 d}} {\langle
    t\rangle^{\frac13}} \|q(0)\|_{\ell^1} \ . \qed
\end{eqnarray*}

\section{Proof of Corollary \ref{NLS}}\label{nonlinear}

Fix any $0<\zeta<\frac13$, $p>5$ and $\theta\in\T^d$. Assume that
  $|V|_r<\varepsilon_{*}$ with $\varepsilon_*$ as in Theorem
  \ref{dispersion_log}. We prove Corollary \ref{NLS} for $t\geq0$, the
  case $t<0$ being totally similar.

First remark that
$
\| H_{\theta} \phi\|_{\ell^\infty}\leq 3 \,
\|\phi\|_{\ell^\infty}
$.
Denote by $f$ the map
$${f:(q_j)_j\mapsto (\mp {\rm i}|q_j|^{p-1} q_j)_j} \ , $$
which describes the nonlinearity in
\eqref{nls}; one has
\begin{align}
  \label{finfty}
\left\|f(q)\right\|_{\ell^\infty}&\leq
\left\|q\right\|_{\ell^{\infty}}^p\ ,
\\
  \label{sti.n.2}
\left\|f(q)\right\|_{\ell^1}&=\sum_{j\in\Z}\left| q_j\right|^p\leq
\left(\sup_{j}\left|q_j\right|^{p-2}\right)\sum_{j}|q_j|^2=\left\|q\right\|_{\ell^{\infty}}^{p-2}\left\|q\right\|_{\ell^2}^2\ .
  \end{align}
In particular, from \eqref{finfty} it follows that \eqref{nls} is
locally well posed in $\ell^{\infty}$. Furthermore,
since the solution of equation \eqref{nls} fulfills
\begin{equation}
  \label{linest2}
\|q(t)\|_{\ell^2} =
\|\phi\|_{\ell^2}\ ,
\end{equation}
it is also globally well posed in $\ell^2$.

Finally we recall the following well known lemma.
\begin{lemma}
  \label{sti.int}
  Let $0<\zeta\leq 1$ and $\mu>1$ be fixed, then $\exists \ C_1>0$ s.t.
  \begin{equation}
    \label{I.conv}
\int_{0}^{t}\frac{1}{\langle
    t-s\rangle^\zeta}\frac{1}{\langle
    s\rangle^{\mu}}ds<\int_{0}^{\infty}\frac{1}{\langle
    t-s\rangle^\zeta}\frac{1}{\langle s\rangle^{\mu}}ds\leq
  \frac{C_1}{\langle t\rangle^{\zeta}}, \quad \forall \ t>0 \ .
    \end{equation}
\end{lemma}

The main step for the proof of the Corollary \ref{NLS} is the next lemma.

{\begin{lemma}
  \label{non.est.ric}
  Define $M:=4K_1$ and $\delta_*:=(C_1 M^{p-2})^{-\frac{1}{p-1}}$. Assume that the initial datum $q(0)=\phi$ for \eqref{nls} fulfills $\delta_0=\left\|\phi\right\|_{\ell^1(\Z)}<\delta_*$, then, if
for some $T>0$ one has
  \begin{equation}
    \label{6}
\sup_{0\leq t\leq T}\langle
t\rangle^{\zeta}\left\|q(t)\right\|_{\ell^\infty} \leq
M\delta_0 \ ,
  \end{equation}
  the solution still fulfills the above inequality with $M$ replaced by
  $\frac{M}2$.
\end{lemma}}
\proof
By Duhamel formula the solution of \eqref{nls} fulfills
  \begin{equation}
    \label{7}
q(t)=e^{-{\rm i} t H_\theta}\phi+\int_0^te^{-{\rm i} (t-s) H_\theta }f(q(s))ds \ .
  \end{equation}
{Under the assumption (\ref{6}), we have, for $0<s\leq T$,
$$  \left\|q(s)\right\|_{\ell^{\infty}}\leq \frac{\delta_0 M}{\la s \ra^{\zeta}} \ .$$
In view of (\ref{sti.n.2}), for $0\leq t \leq T$, the integral is estimated by}
  \begin{align*}
\left\|\int_0^te^{-{\rm i} (t-s) H_\theta }f(q(s))ds
\right\|_{\ell^\infty}&\leq \int_0^t\left\|e^{-{\rm i} (t-s) H_\theta
}f(q(s))\right\|_{\ell^\infty} ds\\
&\leq \int_0^t\frac{K_1}{\langle
  t-s\rangle^{\zeta}}\left\|f(q(s)) \right\|_{\ell^1}ds\\
  &\leq \int_0^t\frac{K_1}{\langle t-s\rangle^{\zeta}}
\left\|q(s)\right\|_{\ell^{\infty}}^{p-2}\left\|q(s)\right\|_{\ell^2}^2ds \\
&\leq \int_0^t\frac{K_1}{\langle t-s\rangle^{\zeta}}\frac{\delta_0^{p-2}M^{p-2}}{\langle
  s\rangle^{\zeta(p-2)}}\left\|\phi\right\|_{\ell^2}^2ds\\
  &=\left\|\phi\right\|_{\ell^2}^2\delta_0^{p-2}M^{p-2} K_1
\int_0^t\frac{1}{\langle t-s\rangle^{\zeta}} \frac{1}{\langle s\rangle^{\zeta(p-2)}}ds\\
&\leq\delta_0^{p}M^{p-2}K_1 \frac{C_1}{\langle
  t\rangle^{\zeta}} \ ,
  \end{align*}
where we used the fact that, under the assumption of the Corollary
  \ref{NLS}, one has $\zeta(p-2)>1$.
Using again \eqref{linest} in order to estimate the term $e^{-i t H_\theta}\phi$ at
r.h.s. of
\eqref{7}, one gets
$$
\sup_{0\leq t\leq T} \left\| q(t)\right\|_{\ell^\infty}\leq \frac{K_1
  \delta_0}{\langle t\rangle^{\zeta}}\left[1+C_1
  M^{p-2}\delta_0^{p-1}\right]\ .
$$
The choice of the constants $M$ and $\delta_*$ made in the statement
of the lemma ensures that the square bracket is smaller than $2$ and
therefore the proof is completed.  \qed

\

\noindent {\it End of the proof of Corollary \ref{NLS}}. First remark
that, by local well-posedness in $\ell^{\infty}$, there exists $T>0$
s.t. \eqref{6} holds. Assume that there exists a finite $T_*$ which is
the largest time for which \eqref{6} holds, then from Lemma
\ref{non.est.ric}, there exists  $T_1>T_*$ s.t. the estimate holds
(the $\ell^{\infty}$ norm
takes some time to move from $\frac{\delta_0 M}{2\langle T_*\rangle^\zeta}$ to $\frac{\delta_0 M}{\langle T_*\rangle^\zeta}$) against the assumption that $T_*$ is the largest time for
which the inequality holds. Thus the solution fulfills \eqref{6} with
$T=\infty$.\qed

\appendix

\section{The fibered rotation number}\label{rotation}

Related to the Schr\"odinger cocycle $(\omega, A_0+F_0)$, we can
define the fibered rotation number $\rho=\rho_{(\omega, A_0+F_0)}$.
It was introduced originally by Herman \cite{Herman} in this
discrete case (see also Johnson-Moser \cite{JM}).
For the precise definition, we follow the same presentation as in \cite{HA}.

Given $ A\in C( \T^d, SL(2,\R))$ with
$A(\cdot)=\left(\begin{array}{cc}
a(\cdot) & b(\cdot) \\
c(\cdot) & d(\cdot)
\end{array}\right)$,
we define the map
$$
\begin{array}{llll}
T_{(\omega,\, A)}:& \displaystyle  \T^d\times \frac12\T &\rightarrow& \displaystyle  \T^d\times \frac12\T \\[3mm]
   &(\theta,\varphi)&\mapsto&(\theta+\omega, \, \phi_{(\omega,\, A)}(\theta,\varphi))
\end{array} \ ,$$
where $\frac12\T:=\R/\pi\Z$ and $\phi_{(\omega,\, A)}(\theta,\varphi)=\arctan\left(\frac{c(\theta)+d(\theta)\tan\varphi}{a(\theta)+b(\theta)\tan\varphi}\right).$
Assume that $A(\theta)$ is homotopic to identity, then the same is true for the map $T_{(\omega,\, A)}$ and therefore it admits a continuous lift
$$
\begin{array}{llll}
\tilde T_{(\omega,\, A)}:& \displaystyle  \T^d\times \R &\rightarrow& \displaystyle  \T^d\times \R \\[2mm]
   &(\theta,\varphi)&\mapsto&(\theta+\omega, \, \tilde\phi_{(\omega,\, A)}(\theta,\varphi))
\end{array}$$
such that $\tilde\phi_{(\omega,\, A)}(\theta,\varphi)\; {\rm mod}\; \pi=\phi_{(\omega,\, A)}(\theta,\varphi\; {\rm mod}\; \pi)$.  The function $$(\theta,\varphi) \mapsto \tilde\phi_{(\omega,\, A)}(\theta,\varphi)-\varphi$$ is $(2\pi)^d-$periodic in $\theta$ and $\pi-$periodic in $\varphi$.
We define now $\rho(\tilde{\phi}_{(\omega,\, A)})$ by
$$\rho(\tilde{\phi}_{(\omega,\, A)})=\limsup_{n\rightarrow+\infty}\frac1n (p_2\circ \tilde T^n_{(\omega,\, A)}(\theta,\varphi)-\varphi )\in\R \ ,$$
where $p_2(\theta,\varphi)=\varphi$.
This limit exists for any $\theta\in\T^d$, $\varphi\in\R$, and the convergence is uniform in $(\theta,\varphi)$ (For the existence of this limit and its properties we can refer to \cite{Herman}).
The class of number $\rho(\tilde\phi_{(\omega,\, A)})$ in $\frac12\T$, independent of the chosen lift, is called the {\bf fibered rotation number} of the skew-product system
$$
\begin{array}{llll}
(\omega, A):& \displaystyle  \T^d\times \R^2 &\rightarrow& \displaystyle  \T^d\times \R^2 \\[2mm]
   &(\theta,\, y)&\mapsto&(\theta+\omega,\, A(\theta)y)
\end{array} \ ,$$
and we denote it by $\rho_{(\omega,\, A)}$.
For further elementary properties, we refer to Appendix of \cite{HA}.

\section{Van der Corput lemma}\label{corput}

\noindent

For the convenience of readers, we give here the statement of the Van
der Corput lemma and its corollary which are used in this paper, even
though they can be found in many textbooks on Harmonic Analysis (see,
e.g., Chapter \uppercase\expandafter{\romannumeral8} of \cite{Stein}).

\begin{Lemma}\label{lemma_VDC}
Suppose that $\psi$ is real-valued and ${\cC}^{k}$ in $(a,b)$ for some $k \geq 2$, and
\begin{equation}\label{lowerbound_derivative}
|\psi^{(k)}(x)| \geq 1,\quad \forall \  x \in (a, b) \ .
\end{equation}
For any $\lambda \in \R^+$, we have
$$\left|\int_a^b e^{{\rm i}\lambda \psi(x)} dx\right|\leq (5\cdot 2^{k-1}-2) \lambda^{-\frac{1}{k}} \ .
$$
\end{Lemma}

If the hypothesis (\ref{lowerbound_derivative}) in the above lemma is replaced by
\begin{equation}\label{lowerbound_derivative_c}
``|\psi^{(k)}(x)| \geq c,\quad \forall \ x \in (a, b)"
\end{equation}
for some $c>0$ independent of $x$, then we can derive from Lemma \ref{lemma_VDC} that
$$\left|\int_a^b e^{{\rm i}\lambda \psi(x)} dx\right|\leq (5\cdot 2^{k-1}-2)c^{-\frac1k} \lambda^{-\frac{1}{k}}, \quad \forall  \  \lambda\in\R_+ \ . $$
Moreover, since (\ref{lowerbound_derivative_c}) also holds for $-\psi$, Lemma \ref{lemma_VDC} implies that
$$\left|\int_a^b e^{{\rm i}\lambda \psi(x)} dx\right|\leq (5\cdot 2^{k-1}-2)c^{-\frac1k} |\lambda|^{-\frac{1}{k}}, \quad \forall  \  \lambda\in\R\setminus\{0\} \ . $$

\begin{Corollary}\label{coro_VDC}
Suppose that $\psi$ is real-valued and ${\cC}^{k}$ in $(a,b)$ for some $k \geq 2$, and that $|\psi^{(k)}(x)| \geq c$ for all $x \in (a, b)$.
Let $h$ be ${\cC}^{1}$ in $(a,b)$. Then
$$\left|\int_a^b e^{{\rm i}\lambda \psi(x)} h(x) dx\right|\leq  (5\cdot 2^{k-1}-2) c^{-\frac1k}\left[|h(b)|+\int_a^b |h'(x)| dx\right] |\lambda|^{-\frac{1}{k}},\quad \forall \ \lambda \in \R\setminus\{0\} \ .$$
\end{Corollary}
This corollary is proved by writing $\displaystyle \int_a^b e^{{\rm i}\lambda \psi(x)}h(x) dx$ as $\displaystyle \int_a^b F'(x)\psi(x) dx$ with
$\displaystyle F(x):=\int_a^x  e^{{\rm i}\lambda \psi(t) } dt$,
integrating by parts, and using the previous estimate
$$|F(x)|\leq  (5\cdot 2^{k-1}-2)c^{-\frac1k} \lambda^{-\frac{1}{k}},\quad \forall \  x\in [a,b] \ .$$

\smallskip

\noindent {\bf Acknowledgments.} The authors would like to thank Hakan Eliasson for fruitful discussions. D. Bambusi acknowledges the support of GNFM.
Z. Zhao is grateful to the support from Laboratoire International Associ\'e (LIA) Laboratoire Ypatia des Sciences Math\'ematiques (LYSM) and from the project MATHIT for traveling to Milan. Z. Zhao would like to thank Chern Institute of Mathematics in Nankai University (Tianjin, China) for the hospitality during his stay.
The authors also appreciate the anonymous referees for helpful suggestions in
modifying this manuscript.

\end{document}